\documentclass[twocolumn,showpacs,preprintnumbers,amsmath,amssymb]{revtex4}

\usepackage{graphicx}
\usepackage{dcolumn}
\usepackage{bm}
\usepackage{textcomp}
\usepackage{wasysym}
\usepackage{textcomp}
\usepackage{mathrsfs}
\usepackage{array}

\begin{document}
\title{Spin-Pair Tunneling in  Mn$_3$ Single-Molecule Magnet}
\author{Yan-Rong Li,  Rui-Yuan Liu and Yun-Ping Wang}
\affiliation{Beijing National Laboratory for Condensed Matter
Physics,  Institute of Physics, Chinese Academy of Sciences, Beijing
100190, People's Republic of China}


\date{\today}
\begin{abstract}
We report spin-pair tunneling observed in Mn$_3$ single-molecule
magnet, which is a crystal with 2D network of identical exchange
coupling. We observed a series of extra quantum tunnelings by the ac
susceptibility measurements, and demonstrated these are concerted
tunnelings of two spins taking place from the same initial state to
the same final state simultaneously. The resonant field of spin-pair
tunneling can be expressed as
$H_z=lD/g\mu_{0}\mu_{B}+(n_{\downarrow}-n_{\uparrow})JS/2{g\mu_{0}\mu_{B}}$,
and the splitting interval ($|J|S/{g\mu_{0}\mu_{B}}$) is half of
that of the single-spin tunneling ($2|J|S/{g\mu_{0}\mu_{B}}$), which
is analogous to the relationship between the magnetic flux quantum
in superconductor ($h/2e$) and common metal ($h/e$).

\end{abstract}
\maketitle
As the counterpart of the electron tunneling, the spin
tunneling in single-molecule magnets (SMMs) manifested by the
quantum tunneling of the magnetization (QTM) has attracted great
interest recently\cite{Mn12Nature,Mn12PRL,FePRL,FeJACS}. In the
existing picture of single-molecule magnets (SMMs), molecules are
highly identical and magnetically independent of each other
 \cite{Mn12Nature,Mn12PRL,FePRL}, hence the spin tunneling of the
molecular clusters doses not rely on its neighbors. The studies on
dimer systems indicate that the intermolecular exchange coupling has
the great influence on the performance of QTM, each half of the
dimer acts as a field bias on its neighbours, shifting the tunneling
resonances to new positions relative to isolated
molecules\cite{dimerNature,dimerPRL,dimerPRB}. Recent research
discovered that for the SMMs with identical exchange coupling (IEC),
the quantum tunneling of molecules heavily depends on its local spin
environments (LSEs)\cite{Mn3identical}. In this letter, we report
spin-pair tunneling (SPT), which represents the concerted quantum
tunneling of two spins taking place from the same initial state to
the same final state simultaneously. The two spins appear to form a
pair and tunnel as a unit. SPT is clearly identified in the ac
susceptibility curves of Mn$_3$ SMM, and evidenced by the abnormally
high effective barrier at zero field as well.

The crystal of Mn$_{3}$ SMM has the formula
[Mn${_3}$O(Et-sao)${_3}$(MeOH)${_3}$(ClO${_4}$)]. The preparation
and crystal characteristics of Mn$_{3}$ SMM have been reported in
earlier literatures \cite{Mn3distort, Mn3identical}.  As described
in Ref.\cite{Mn3identical}, Mn$_3$ SMM is a crystal with 2D network
of exchange coupling, in which each molecule is coupled with three
neighboring molecules by hydrogen bonds in ab plain, forming a
honeycomb-like structure viewed down along the c-axis, therefore
Mn$_3$ SMM are considered to be a crystal with IEC and a model
systems of simple Ising model. Each molecule has the ground spin
state of $S=6$ and a spin Hamiltonian of $\hat{H}=-D\hat{S}_z{}^2 +
g\mu_{0}\mu_{B}\hat{S}_zH_z$, where $D=0.98$K, $g=2{_\cdot06}$
\cite{Mn3distort}. Due to the identical exchange coupling in Mn$_3$
SMM, the quantum tunneling is equally split in the way of
$(n_{\downarrow}-n_{\uparrow})JS/{g\mu_{0}\mu_{B}}$, where
$n_{\downarrow}$ and $n_{\uparrow}$ represent the numbers of
spin-down and spin-up  molecules neighboring to the tunneling
molecule, and $J=-0.041$K is the intermolecular exchange coupling
constant\cite{Mn3identical}.

Below the blocking temperature, SMMs show slow magnetic relaxation
as spin-flippings become difficult due to the high energy barrier,
whereas SMMs show fast magnetic relaxation  at the resonant
tunneling field because of quantum tunneling effect, which leads to
the step-like hysteresis
loops\cite{Mn12Nature,Mn12PRL,FePRL,FeJACS}.
 Apart from dc susceptibility measurement, ac susceptibility measurement is also considered a good way to
 define the magnetic relaxation behavior in SMMs.
 Since the magnetic
 relaxation time obviously decreases at the resonant tunneling field, ac
 susceptibility  demonstrates the peaks at the resonant tunneling
 fields\cite{ac susceptibility}.

 The blocking temperature of Mn$_3$
SMM is estimated to be 3K \cite{Mn3identical}, hence we measured ac
susceptibility at temperatures above 3k. Fig.1(a) shows the field
dependence of ac susceptibility at 7K with a frequency of 9.99KHz.
\begin{figure}[ht] \scalebox{0.56}{\includegraphics[bb=380 3 8cm
22.5cm]{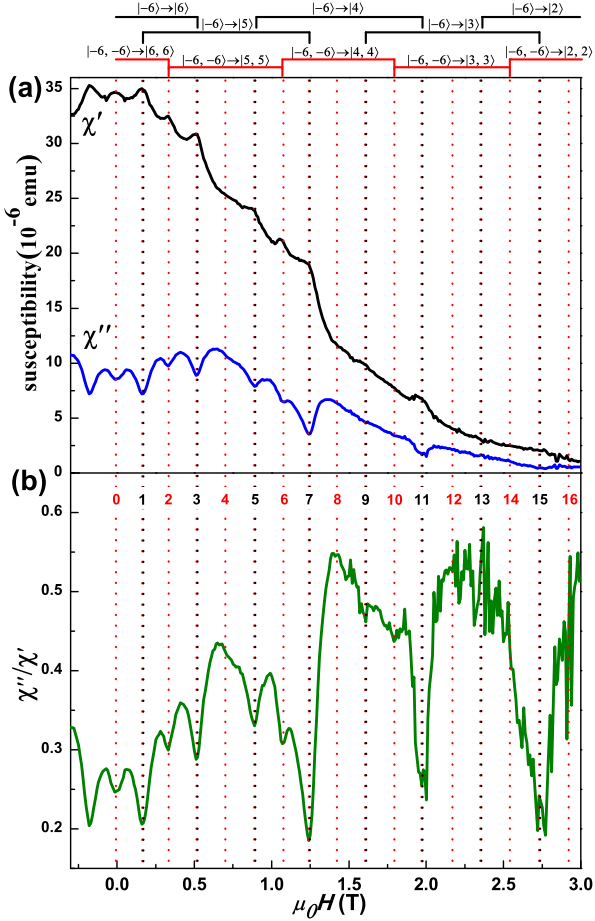}}
 \caption
{(Color online). (a) Field dependence of ac susceptibility $\chi{'}$
(real component) and $\chi{''}$ (imaginary component) from $-$0.3T
to 3T at 7K, with the sweeping rate of 0.001T/s and frequency of
9.99KHz. (b) Field dependence of $\chi{'}$ /$\chi{''}$ from $-$0.3T
to 3T. The quantum tunnelings marked by black and red dotted lines
are single-spin tunnelings and spin-pair tunnelings in different
quantum tunneling sets, respectively \cite{Supplemental Material}.}
\end{figure}
A series of peaks and dips have been observed in $\chi{'}$ (real
component) and $\chi{''}$(imaginary component) curve respectively.
Field dependence of $\chi{''}/\chi{'}$ is shown in Fig.1(b), which
is considered a quantity proportional to the relaxation
times\cite{spin dynamic}, clearly $\chi{''}/\chi{'}$ demonstrates
dips at the resonant fields. Apparently, in addition to the quantum
tunnelings numbered as 1, 3, 5, 7, 9, 11, 15, which are due to the
spin tunneling of single Mn$_3$ molecule in different spin
environments according to Ref.\cite{Mn3identical}, four extra
quantum tunnelings numbered as 0, 2, 6, 10, and located at 0T,
0.34T, 1.08T, 1.80T respectively are observed.  Here the location of
all the resonant fields are temperature and frequency independent.
Phenomenally, each of the extra quantum tunneling mentioned above
happens to appear at the midpoint between its two neighbor
tunnelings. It is also noticeable that, there is a quantum tunneling
taking place at zero field. It had been suspected that a different
type of isolated molecule might exist, which contributed to the
extra tunnelings. However, this conjecture is dismissed, since the
four-circle diffraction measurement shows the sample is a good
single crystal, and Fig.1 shows that every resonant field of the
extra tunnelings is located at the midpoint between its two
neighboring tunnelings. We will demonstrate in the following that,
these extra tunnelings as well as the tunneling occurring at zero
field are of spin-pair tunnelings (SPT), i. e. concerted tunneling
by two spins taking place from the same initial state to the same
final state simultaneously.

In Mn$_3$ SMM, a single molecule has three exchange-coupled
neighbors, hence,  a pair of Mn$_3$ molecules has four
exchange-coupled neighbors. Fig.2 demonstrates the five local spin
environments (LSEs) of a spin pair marked in black, which is
labelled as $(n_{\downarrow},n_{\uparrow})$ as described in
Ref.\cite{Mn3identical}.

\begin{figure}[ht] \scalebox{0.73} {\includegraphics[bb=30 0 8cm
12cm]{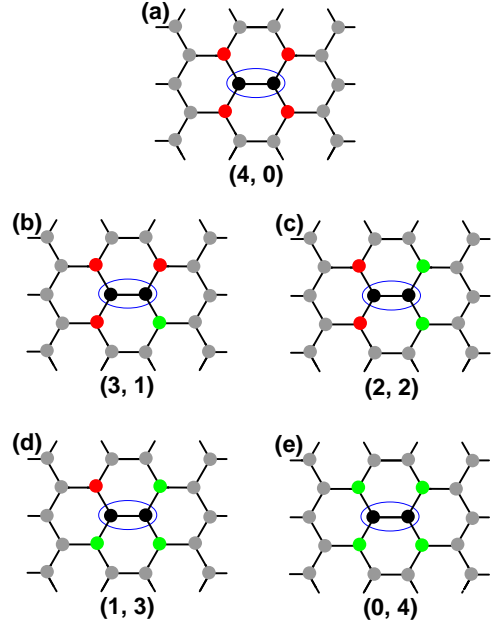}}
 \caption
{(Color online).Sketch maps of five spin configurations with
different LSE $(n_{\downarrow},n_{\uparrow})$ for a pair of
molecules, other equivalent spin configurations are not listed here
for simplicity. The tunneling pair in blue ellipse is marked in
black color, which could occupy either spin-up or spin-down state
simultaneously. Its four neighboring molecules with red and green
color occupy spin-down and spin-up states respectively. The
direction of spin is perpendicular to the honeycomb lattice of
Mn$_3$ SMM. The black lines between molecules represent the exchange
couplings. }
\end{figure}
For the SMMs with IEC, the spin Hamiltonian of each molecule may be
presented as:
\begin{equation}
\hat{\mathcal{H}}=-D\hat{S}^2_z +
g\mu_{0}\mu_{B}\hat{S}_zH_z-\sum_{i=1}^{n}J\hat{S}_{z}\hat{S}_{iz}+\hat{\mathcal{H}}^{trans},
\end{equation}
where $n$ is coordination number, $\hat{S}_{z}$ and $\hat{S}_{iz}$
are the easy-axis spin operators of the molecule and its $i$th
exchange-coupled neighboring molecule, $\hat{\mathcal{H}}^{trans}$
is the small off-diagonal perturbation term which allows the quantum
tunneling to occur\cite{Mn3identical}. For the single-spin tunneling
(SST) from $|-S\rangle$ to $|S-l\rangle$ ($l=0,1,2,3,....$), the
resonant field is determined by
 \begin{equation}
H_z=lD/g\mu_{0}\mu_{B}+(n_{\downarrow}-n_{\uparrow})JS/{g\mu_{0}\mu_{B}},
\end{equation}
and hence the quantum tunneling from $|-6\rangle$ to $|6\rangle$
spin state is split into four, which occurs at
$3JS/{g\mu_{0}\mu_{B}}$, $JS/{g\mu_{0}\mu_{B}}$,
$-JS/{g\mu_{0}\mu_{B}}$, $-3JS/{g\mu_{0}\mu_{B}}$
respectively\cite{Mn3identical}.  However, for the SPT from
$|-S,-S\rangle$ to $|S-l,S-l\rangle$ (the spin state of the pair can
be represented as $|m,m\rangle$, where m is the spin quantum number
of the two spins in one pair),  the resonant field is determined by
 \begin{equation}
H_z=lD/g\mu_{0}\mu_{B}+(n_{\downarrow}-n_{\uparrow})JS/2{g\mu_{0}\mu_{B}},
\end{equation}
and hence the quantum tunneling from $|-6, -6\rangle$ to $|6,
6\rangle$ spin state is split into five, which occurs at
$2JS/{g\mu_{0}\mu_{B}}$, $JS/{g\mu_{0}\mu_{B}}$, 0,
$-JS/{g\mu_{0}\mu_{B}}$, $-2JS/{g\mu_{0}\mu_{B}}$ respectively.
Apparently the splitting interval of SPT is
$|J|S/{g\mu_{0}\mu_{B}}$, which is half of that of SST.

Fig.1 has demonstrated all SSTs and SPTs in different quantum
tunneling sets \cite{Supplemental Material}, which are marked by
black and red dotted lines, respectively. Within a SST set, the LSEs
are (3, 0), (2, 1), (1, 2), (0, 3) from the left to the right,
respectively, and within a SPT set, the LSEs are (4, 0), (3, 1), (2,
2), (1, 3), (0, 4) from the left to the right, respectively.

Since the axial anisotropy constant $D=0.98$K happens to be close to
$4|J|S$ in Mn$_3$ SMM \cite{Mn3FM,Mn3identical}, there are quantum
tunnelings overlapping, for example, the quantum tunneling numbered
as 1 is the combinations of the SST from $|-6\rangle$ to $|6\rangle$
spin state with LSE (1, 2) and the SST from $|-6\rangle$ to
$|5\rangle$ spin state with LSE (3, 0) and the SPT from $|-6,
-6\rangle$ to $|6, 6\rangle$ spin state with LSE (1, 3). With such
coincidence, the quantum tunnelings are taking place at fields with
equal interval, just as that happens to the individual
SMMs\cite{Mn12PRL, FeJACS}. Of the overlapped tunnelings mentioned
above, the contribution of the component quantum tunnelings are
different due to the dependence of tunneling on the local spin
environment and the potential barrier. It is noticed that the extra
quantum tunnelings numbered with even numbers are purely SPT,  while
the quantum tunnelings numbered with odd numbers are of the
combination of SST and SPT.
 As described in Ref.\cite{Mn3identical}: the tunneling
\begin{figure}[hb]
\scalebox{0.48}{\includegraphics[bb=140 10 8cm 20.2cm]{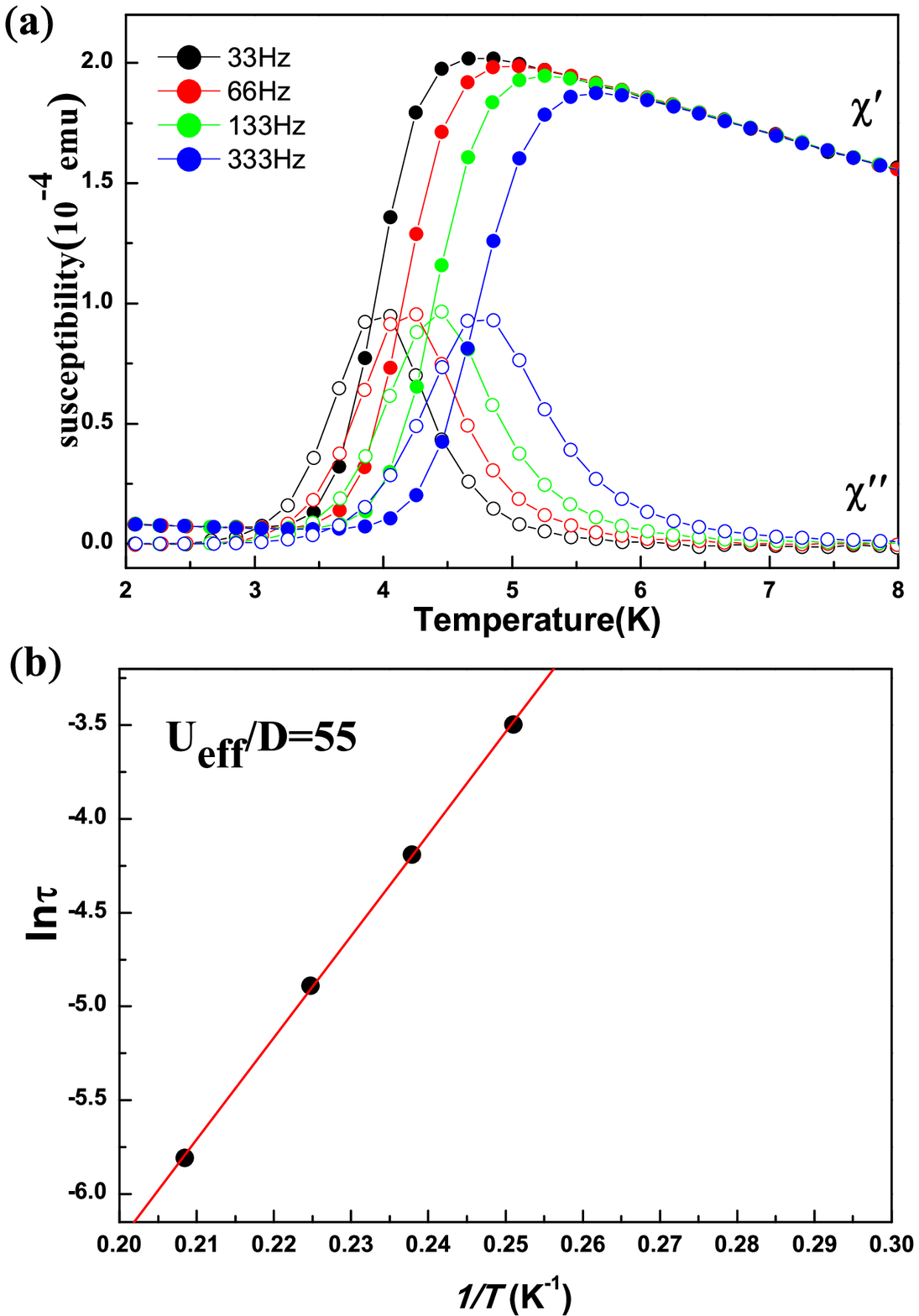}}
\quad \scalebox{0.33}{\includegraphics[bb=36 -52 8cm
4cm]{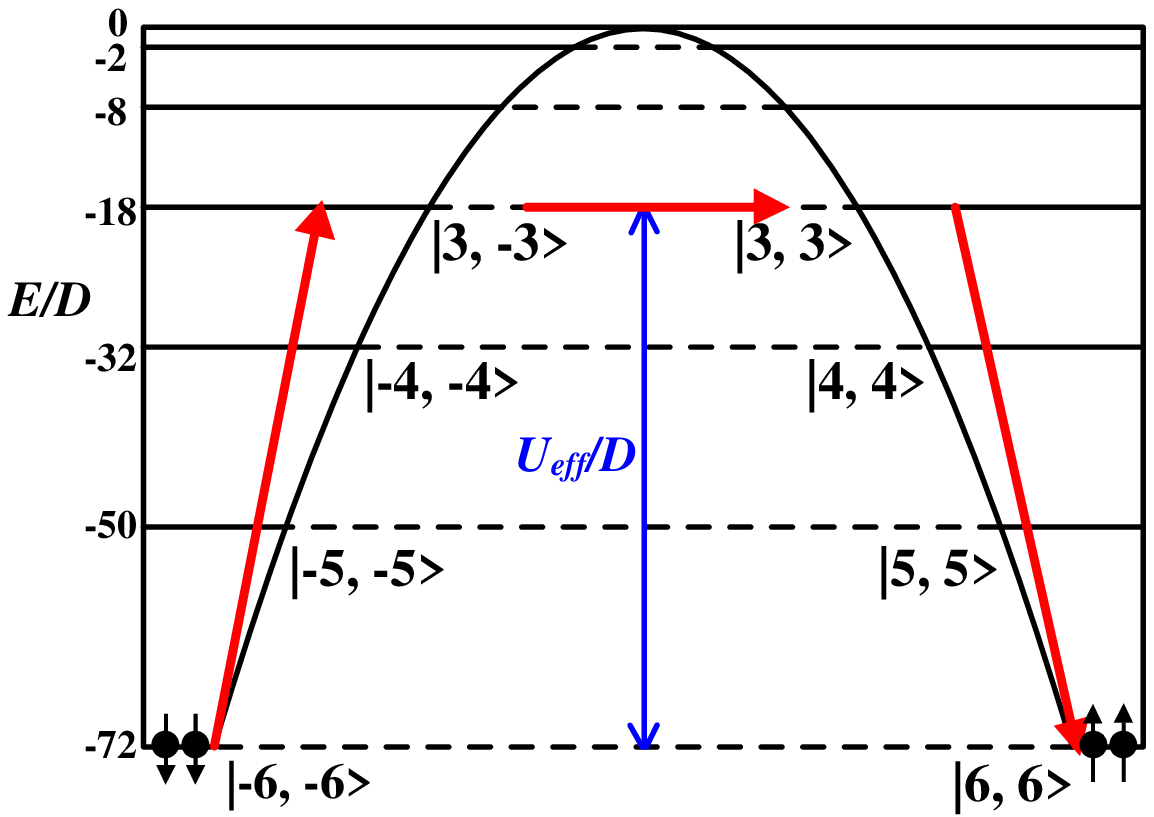}}
 \caption {(Color online). (a) Temperature dependence of ac susceptibility
 at different frequencies. (b) plot of $ln(\tau)$ vs $1/T$ with data obtained from ac susceptibility
 measurement, which gives $U_{eff}/D=55$,
 the inset shows the schematic drawing for the magnetic relaxation process within temperature range 4$\sim$5K. }
\end{figure}
 magnitude $\mathcal{T}$ of SST is described as:
\begin{equation} \mathcal{T}=\alpha
  N_{(n_{\downarrow},n_{\uparrow})}
  P_{|m_{i}\rangle\rightarrow|m_{f}\rangle},
  \end{equation}
where $N_{(n_{\downarrow}, n_{\uparrow})}$ is the number of
molecules with LSE $(n_{\downarrow},n_{\uparrow})$,
$P_{|m_{i}\rangle\rightarrow|m_{f}\rangle}$ is the tunneling
probability of the molecule from the spin state $|m_{i}\rangle$ to
$|m_{f}\rangle$, which is exponentially dependent on the effective
barrier at the thermally activated tunneling region\cite{Li,ac
susceptibility}. Note that Eq.(4) is applicable to SPT with the
following transformation:
\begin{equation} \mathcal{T}=\alpha
N_{(n_{\downarrow},n_{\uparrow})} P_{|m_{i},
m_{i}\rangle\rightarrow|m_{f}, m_{f}\rangle},
\end{equation}
where $N_{(n_{\downarrow}, n_{\uparrow})}$ is the number of spin
pairs with the LSE $(n_{\downarrow}, n_{\uparrow})$, and $P_{|m_{i},
m_{i}\rangle\rightarrow|m_{f}, m_{f}\rangle}$ is determined by the
aggregated effective barrier of two SSTs, which can be deduced from
$P_{|m_{i}, m_{i}\rangle\rightarrow|m_{f},
m_{f}\rangle}\propto{P_{|m_{i}\rangle\rightarrow|m_{f}\rangle}^2}$
and
$P_{|m_{i}\rangle\rightarrow|m_{f}\rangle}\propto{}\exp{(-U_{eff}/k_{B}T)}$.
Apparently, the effective barrier of SPT is doubled, and hence much
higher than that of SST, therefore, the tunneling magnitude of SPT
is much smaller than SST, and it is of no surprise that only the
quantum tunnelings of SST are observed in hysteresis loop at
2K\cite{Mn3identical}. According to Eq. (4) and (5), the tunneling
magnitudes of SST and SPT are heavily dependent on the numbers of
single spins and spin pairs in the proper LSEs respectively. At a
high positive field, most molecules occupy $|6\rangle$ spin state,
therefore SSTs with LSE (0, 3) are phenomenal, such as the quantum
tunnelings numbered as 7, 11, 15 observed in Fig.1. On the other
hand, only SPTs with LSE (0, 4) are observed at high field, such as
the quantum tunnelings numbered as 6, 10 in Fig.1, and the expected
quantum tunnelings numbered as 4, 8, 12, 13, 16 are not observed due
to very small $N_{(n_{\downarrow},n_{\uparrow})}$, whereas the
absence of SPT numbed as 14 is due to small signal-to-noise ratio.
\begin{table*}[!t]
\caption{The similarities and differences between spin-pair
tunneling and Josephson effect. }
\begin{center}
\renewcommand\arraystretch{1.25}
\begin{tabular}{|c|c|c|}
\hline
  \hspace{2mm}\hspace{2mm}&  \hspace{14mm} \textbf{spin-pair tunneling}\hspace{14mm} & \hspace{3mm} \textbf{Josephson effect} \hspace{3mm}    \\
\hline \textbf{tunneling unit} &  spin pair & Cooper pair \\
\hline
 \textbf{pairing partner }& not fixed   &  not fixed   \\
\hline
 \textbf{internal state of pair} &  exchange-energy entanglement state   & momentum and spin entanglement state\\
    \hline
    \textbf{
invariant of pair} & net exchange energy is a constant  & both net momentum and net spin are zero  \\
    \hline
     \textbf{tunneling variable} & spin orientation of spin pair  & position of Cooper pair  \\
      \hline
\textbf{relationship}&splitting interval $|J|S/{g\mu_{0}\mu_{B}}$ is half of & quantum flux $h/2e$ is half of\\
  \textbf{with single particle}&that of the single-spin tunneling ($2|J|S/{g\mu_{0}\mu_{B}}$)&that in common metal ($h/e$)\\
    \hline

\end{tabular}
\end{center}
\end{table*}

Fig.3(a) shows ac susceptibility of Mn$_3$ as a function of
temperature with different frequencies at zero field, which
demonstrates typical characteristic of SMMs: The dissipation peak in
$\chi{''}$-$T$ curve drops to lower temperature at higher
frequency\cite{Mn12 relaxation, acFe8, TeJACS}. Fig.3(b) shows the
fitting of the Arrhenius equation, which gives the effective barrier
$U_{eff}=55$K, in good agreement with the result mentioned in
Ref\cite{Mn3distort}. It is remarkable that $U_{eff}=55$K is much
larger than the anisotropy barrier $DS_{z}^2=35$K, whereas $U_{eff}$
is usually smaller than $DS_{z}^2$ in the previous SMM
studies\cite{Li,ac susceptibility}.  As mentioned above, the
effective barrier of SPT is supposed to be the double of that of
SST, and since $U_{eff}=55$K happens to be close to the double of
the energy gap (26K) between $|\pm6\rangle$ and $|\pm3\rangle$. It
is evident that  SPT leads to the observed tunneling at zero field,
i. e.  the spin relaxation at zero field should be dominated by two
concerted spin flippings, and each flipping is of thermally assisted
tunneling\cite{ac susceptibility}. As shown in the inset of
Fig.3(b), the spin pair is initially thermally activated from $|-6,
-6\rangle$ state to $|-3, -3\rangle$ state, next tunnels to $|3,
3\rangle$ state, and finally relaxes to $|6, 6\rangle$ state, this
is similar to the relaxation process in the previous SMMs
studies\cite{Li, ac susceptibility}. Note that quantum tunneling
taking place from $|-3, -3\rangle$ to $|3, 3\rangle$ state in this
relaxation process is consistent with the selection rule of C$_3$
symmetry in Mn$_3$ SMM ($\Delta{m_{s}}=3n$) \cite{selection}.

SPT in Mn$_3$ SMM  is analogous to the tunneling of superconducting
Cooper pairs, i.e. Josephson effect in superconductor, in the sense
of that, the two spins of a spin pair entangle to behave as a unit,
while the pairing could be formed between a spin and any rather than
a particular one of its neighboring spins.  It is well known that
both the net momentum and spin of the Cooper pairs are zero
regardless of the value of the individual momentum and spin, which
can be considered as a source of momentum and spin entanglement
state\cite{BCS,spin-entanglment}. Similarly, the net exchange energy
of spin pair is a constant regardless of the individual exchange
energy distribution, for example, the SPT at zero field requires the
LSE of spin pairs to be (2, 2), which means the net exchange
interaction between the spin pair and its neighbors equals to zero,
nevertheless there are six equivalent spin distributions (Fig.2 (c)
only shows one of them). Furthermore, the splitting interval of SPT
is $|J|S/{g\mu_{0}\mu_{B}}$ according to Eq. (3), which is half of
that of SST ($2|J|S/g\mu_{0}\mu_{B}$), this is similar to the
relationship between the magnetic flux quantum in superconductor(
$h/2e$) and common metal ($h/e$). The similarities and differences
between SPT and Josephson effect are shown in Table.1. It is notable
that spin-pair tunneling is attributed to the identical exchange
coupling between the molecules, and hence cannot be observed in the
SMMs without exchange coupling such as Mn$_{12}$\cite{Mn12PRL,ac
susceptibility}, Fe$_8$\cite{FePRL}, etc. To the best of our
knowledge, this is the first time that the spin-pair tunneling is
observed and reported, which not only adds to the diversity of QTM,
but also opens up new perspectives in the quantum physics and
potential applications of these molecular nanomagnets.

 This work was supported by the National Key Basic Research Program of China
(No.2011CB921702).


\begin{thebibliography}{21}
\bibitem{Mn12PRL} J. R. Friedman, M. P. Sarachik, J. Tejada, and R. Ziolo,  Phys. Rev. Lett. \textbf{76}, 3830 (1996).
\bibitem{Mn12Nature} L. Thomas, F. Lionti, R. Ballou, D. Gatteschi, R. Sessoli, and B. Barbara,  Nature \textbf{383}, 145 (1996).
\bibitem{FePRL} C. Sangregorio, T. Ohm, C. Paulsen, R. Sessoli, and D. Gatteschi, Phys. Rev. Lett. \textbf{78}, 4645 (1997).
\bibitem{FeJACS} K. L. Taft, C. D. Delfs, G. C. Papaefthymiou, S. Foner, D. Gatteschi, and S. J. Lippard, J. Am. Chem. Soc. \textbf{116}, 823 (1994).
\bibitem{dimerNature} W. Wernsdorfer, N. Aliaga-Alcalde, D. N. Hendrickson, and G. Christou, Nature \textbf{416}, 406 (2002).
\bibitem{dimerPRL} W. Wernsdorfer, S. Bhaduri, R. Tiron, D. N. Hendrickson, and G. Christou, Phys. Rev. Lett. \textbf{89} 197201 (2002).
\bibitem{dimerPRB} R. Tiron, W. Wernsdorfer, N. Aliaga-Alcalde, and G. Christou, Phys. Rev. B \textbf{68} 140407(R) (2003).
\bibitem{Mn3distort} R. Inglis, L. F. Jones, G. Karotsis, A. Collins, S. Parsons, S. P. Perlepes, W. Wernsdorfer, and E. K. Brechin, Chem. Commun., 5924 (2008).
\bibitem{Mn3identical} Y. R. Li, R. Y. Liu, H. Q. Liu, and Y. P. Wang,  Phys. Rev. B  \textbf{89}, 184401 (2014).
\bibitem{ac susceptibility} F. Luis, J. Bartolome, J. F. Fernandez, J. Tejada, J. M. Hernandez, X. X. Zhang, and R. Ziolo, Phys. Rev. B \textbf{55}, 11448
(1997).
\bibitem{spin dynamic} T. Pohjola, H. Schoeller,  Phys. Rev. B \textbf{62}, 15026 (2000).
\bibitem{Supplemental Material} See Supplemental Material at[] for the detailed
 description of the expected quantum tunnelings labeled from 0 to
16 in Fig.1.
\bibitem{Mn3FM}R. Inglis, S. M. Taylor, L. F. Jones, G. S. Papaefstathiou, S. P. Perlepes, S. Datta, S. Hill, W. Wernsdorfer, and E. K. Brechin, Dalton Trans. 9157 (2009).
\bibitem{Li} Y. R. Li, H. Q. Liu, Y. Liu, S. K. Su, and Y. P. Wang,  Chin. Phys. Lett. \textbf{26}, 077504 (2009).
\bibitem{Mn12 relaxation} A. M. Gomes, M. A. Novak, R. Sessoli, A. Caneschi, and D. Gatteschi,
Phys. Rev. B \textbf{57}, 5021 (1998).
\bibitem{acFe8} X. X. Zhang, J. M. Hernandez, E. del Barco, J. Tejada, A. Roig, E. Molins, and K. Wieghardt, J. Appl. Phys. \textbf{85}, 5633
(1999).
\bibitem{TeJACS} S. Accorsi, et al., J. Am. Chem. Soc. \textbf{128}, 4742 (2006).
\bibitem{selection} J. J. Henderson, C. Koo, P. L. Feng, E. del Barco, S. Hill, I. S. Tupitsyn, P. C. E. Stamp, and D. N. Hendrickson, Phys. Rev. Lett. \textbf{103}, 017202 (2009).
\bibitem{BCS} J. Bardeen, L. N. Cooper, and J. R. Schrieffer, Phys. Rev. \textbf{108}, 1175 (1957).
\bibitem{spin-entanglment} A. Hayat, H-Y. Kee, K. S. Burch, and A. M. Steinberg, Phys. Rev. B \textbf{89}, 094508 (2014).
 \end{thebibliography}
\end{document}


\title{Supplemental Material to ``Spin-Pair Tunneling in  Mn$_3$ Single-Molecule Magnet"}
\author{Yan-Rong Li,  Rui-Yuan Liu and Yun-Ping Wang}
\affiliation{Beijing National Laboratory for Condensed Matter
Physics,  Institute of Physics, Chinese Academy of Sciences, Beijing
100190, People's Republic of China}

\maketitle We here present the detailed description of the expected
quantum tunnelings, which are labeled from 0 to 16 in Fig.1. Table.1
lists
\begin{table}
\caption{ The detailed description of the expected quantum
tunnelings labeled from 0 to 16 in Fig.1.}
\begin{center}
\renewcommand\arraystretch{1.1}
\begin{tabular}{|c|c|c|c|c|}
\hline
 $ \hspace{3mm}\textbf{n}^{\circ}$\hspace{3mm}&  \hspace{12mm}$|i\rangle$$\Longleftrightarrow$$|f\rangle$\hspace{12mm} & \hspace{1mm} LSE($n_{\downarrow}, n_{\uparrow})$\hspace{1mm}  &\hspace{3mm}Type\hspace{3mm} &\hspace{3mm}Resonant Field\hspace{3mm}  \\
\hline \textbf{0} & $|-6,-6\rangle$$\Longleftrightarrow$$|6,6\rangle$ &(2, 2)& SPT &0 \\
\hline   &$|-6\rangle$$\Longleftrightarrow$$|6\rangle$  & (1, 2)& SST&$|J|S/{g\mu_{0}\mu_{B}}$ \\
 \textbf{1} &$|-6\rangle$$\Longleftrightarrow$$|5\rangle$  & (3, 0)& SST&$(D-3|J|S)/{g\mu_{0}\mu_{B}}$\\
    &$|-6,-6\rangle$$\Longleftrightarrow$$|6,6\rangle$  & (1, 3)& SPT&$|J|S/{g\mu_{0}\mu_{B}}$\\
    \hline
     \textbf{2} &$|-6,-6\rangle$$\Longleftrightarrow$$|6,6\rangle$ &(0, 4)&SPT&$2|J|S/{g\mu_{0}\mu_{B}}$\\
 &$|-6,-6\rangle$$\Longleftrightarrow$$|5,5\rangle$ &(4, 0)&SPT &$(D-2|J|S)/{g\mu_{0}\mu_{B}}$\\
     \hline &$|-6\rangle$$\Longleftrightarrow$$|6\rangle$  & (0, 3)& SST&$3|J|S/{g\mu_{0}\mu_{B}}$\\
 \textbf{3} &$|-6\rangle$$\Longleftrightarrow$$|5\rangle$  & (2, 1)& SST&$(D-|J|S)/{g\mu_{0}\mu_{B}}$\\
     &$|-6,-6\rangle$$\Longleftrightarrow$$|5,5\rangle$  & (3, 1)& SPT&$3|J|S/{g\mu_{0}\mu_{B}}$\\
\hline
     \textbf{4}
 &$|-6,-6\rangle$$\Longleftrightarrow$$|5,5\rangle$ &(2, 2)&SPT &$D/{g\mu_{0}\mu_{B}}$\\
    \hline
    &$|-6\rangle$$\Longleftrightarrow$$|5\rangle$  & (1, 2)& SST&$(D+|J|S)/{g\mu_{0}\mu_{B}}$\\
 \textbf{5} &$|-6\rangle$$\Longleftrightarrow$$|4\rangle$  & (3, 0)& SST&$(2D-3|J|S)/{g\mu_{0}\mu_{B}}$\\
     &$|-6,-6\rangle$$\Longleftrightarrow$$|5,5\rangle$  & (1, 3)& SPT&$(D+|J|S)/{g\mu_{0}\mu_{B}}$ \\
    \hline
     \textbf{6} &$|-6,-6\rangle$$\Longleftrightarrow$$|5,5\rangle$ &(0, 4)&SPT &$(D+2|J|S)/{g\mu_{0}\mu_{B}}$\\
&$|-6,-6\rangle$$\Longleftrightarrow$$|4,4\rangle$ &(4, 0)&SPT &$(2D-2|J|S)/{g\mu_{0}\mu_{B}}$\\
     \hline
      &$|-6\rangle$$\Longleftrightarrow$$|5\rangle$  & (0, 3)& SST&$(D+3|J|S)/{g\mu_{0}\mu_{B}}$\\
 \textbf{7} &$|-6\rangle$$\Longleftrightarrow$$|4\rangle$  & (2, 1)& SST&$(2D-|J|S)/{g\mu_{0}\mu_{B}}$ \\
    &$|-6,-6\rangle$$\Longleftrightarrow$$|4,4\rangle$  & (3, 1)& SPT&$(2D-|J|S)/{g\mu_{0}\mu_{B}}$\\
\hline
     \textbf{8}  &$|-6,-6\rangle$$\Longleftrightarrow$$|4,4\rangle$ &(2, 2)&SPT&$2D/{g\mu_{0}\mu_{B}}$\\
    \hline
     &$|-6\rangle$$\Longleftrightarrow$$|4\rangle$  & (1, 2)& SST&$(2D+|J|S)/{g\mu_{0}\mu_{B}}$\\
 \textbf{9} &$|-6\rangle$$\Longleftrightarrow$$|3\rangle$  & (3, 0)& SST&$(3D-3|J|S)/{g\mu_{0}\mu_{B}}$\\
    &$|-6,-6\rangle$$\Longleftrightarrow$$|4,4\rangle$  & (1, 3)& SPT&$(2D+|J|S)/{g\mu_{0}\mu_{B}}$\\
    \hline
   \textbf{10} &$|-6,-6\rangle$$\Longleftrightarrow$$|4,4\rangle$ &(0, 4)&SPT&$(2D+2|J|S)/{g\mu_{0}\mu_{B}}$\\
&$|-6,-6\rangle$$\Longleftrightarrow$$|3,3\rangle$ &(4, 0)&SPT&$(3D-2|J|S)/{g\mu_{0}\mu_{B}}$\\
     \hline
      &$|-6\rangle$$\Longleftrightarrow$$|4\rangle$  & (0, 3)& SST&$(2D+3|J|S)/{g\mu_{0}\mu_{B}}$\\
 \textbf{11} &$|-6\rangle$$\Longleftrightarrow$$|3\rangle$  & (2, 1)& SST&$(3D-|J|S)/{g\mu_{0}\mu_{B}}$\\
    &$|-6,-6\rangle$$\Longleftrightarrow$$|3,3\rangle$  & (3, 1)& SPT&$(3D-|J|S)/{g\mu_{0}\mu_{B}}$\\
\hline
     \textbf{12}  &$|-6,-6\rangle$$\Longleftrightarrow$$|3,3\rangle$ &(2, 2)&SPT&$3D/{g\mu_{0}\mu_{B}}$\\
\hline
     &$|-6\rangle$$\Longleftrightarrow$$|3\rangle$  & (1, 2)& SST&$(3D+|J|S)/{g\mu_{0}\mu_{B}}$\\
 \textbf{13} &$|-6\rangle$$\Longleftrightarrow$$|2\rangle$  & (3, 0)& SST&$(4D-3|J|S)/{g\mu_{0}\mu_{B}}$\\
    &$|-6,-6\rangle$$\Longleftrightarrow$$|3,3\rangle$  & (1, 3)& SPT&$(3D+|J|S)/{g\mu_{0}\mu_{B}}$\\
\hline
     \textbf{14} &$|-6,-6\rangle$$\Longleftrightarrow$$|3,3\rangle$ &(0, 4)&SPT&$(3D+2|J|S)/{g\mu_{0}\mu_{B}}$\\
&$|-6,-6\rangle$$\Longleftrightarrow$$|2,2\rangle$ &(4, 0)&SPT&$(4D-2|J|S)/{g\mu_{0}\mu_{B}}$\\
    \hline
     &$|-6\rangle$$\Longleftrightarrow$$|3\rangle$  & (0, 3)& SST&$(3D+3|J|S)/{g\mu_{0}\mu_{B}}$\\
 \textbf{15} &$|-6\rangle$$\Longleftrightarrow$$|2\rangle$  & (2, 1)& SST&$(4D-|J|S)/{g\mu_{0}\mu_{B}}$\\
    &$|-6,-6\rangle$$\Longleftrightarrow$$|2,2\rangle$  & (3, 1)& SPT&$(4D-|J|S)/{g\mu_{0}\mu_{B}}$\\
        \hline
     \textbf{16}  &$|-6,-6\rangle$$\Longleftrightarrow$$|2,2\rangle$ &(2, 2)&SPT& $4D/{g\mu_{0}\mu_{B}}$ \\
     \hline
\end{tabular}
\end{center}
\end{table}
the initial state $|i\rangle$ and the final state $|f\rangle$, the
local spin environment (LSE), and tunneling type, and theoretical
calculation of resonant field for the quantum tunnelings.
$|m\rangle$ and $|m, m\rangle$ stand for the spin states of one
single spin and spin pair, where m is the spin quantum number. LSE
is labbeld as ($n_{\downarrow}, n_{\uparrow})$, where
$n_{\downarrow}$ and $n_{\uparrow}$ represent the numbers of
spin-down and spin-up spins neighboring to the tunneling spins or
spin pairs. SPT and SST are short forms for spin-pair tunneling and
single-spin tunneling respectively.  $D=0.98$K is the axial
anisotropy constant, $J=-0.041$K is the intermolecular exchange
coupling constant.